\documentclass[structabstract]{aa}
\usepackage{graphicx}
\usepackage{amsmath}  
\usepackage{enumerate} 
\usepackage{txfonts}
\usepackage{natbib}
\bibpunct{(}{)}{;}{a}{}{,} 

\begin{document}


\title{Ionization toward the high-mass star-forming region NGC~6334~I\thanks{{\it Herschel} is an ESA space observatory with science instruments provided by European-led Principal Investigator consortia and with important participation from NASA.}}

\author{Jorge L. Morales Ortiz\inst{\ref{inst1},\ref{inst2}} \and
               Cecilia Ceccarelli\inst{\ref{inst1}} \and
               Dariusz C. Lis\inst{\ref{inst3}} \and
               Luca Olmi\inst{\ref{inst2},\ref{inst4}} \and
               Ren\'e Plume\inst{\ref{inst5}} \and
               Peter Schilke\inst{\ref{inst6}} }

\institute{
UJF-Grenoble 1 / CNRS-INSU, Institut de Plan\'etologie et d'Astrophysique de Grenoble
(IPAG) UMR 5274, Grenoble, F-38041, France \email{jorge.luis379@gmail.com} \label{inst1}
\and
University of Puerto Rico, R\'{i}o Piedras Campus, Physics Department, Box 23343,
UPR station, San Juan, Puerto Rico (USA) \label{inst2}
\and
California Institute of Technology, Pasadena, CA 91125, USA \label{inst3}
\and
Osservatorio Astrofisico di Arcetri - INAF, Largo E. Fermi 5, I-50125, Firenze, Italy \label{inst4}
\and
Department of Physics and Astronomy, University of Calgary, Calgary, AB T2N 1N4, Canada \label{inst5}
\and
I. Physikalisches Institut der Universit\"{a}t zu K\"{o}ln, Z\"{u}lpicher Str. 77, 50937 K\"{o}ln, Germany \label{inst6} }

\date{Received; accepted }

\abstract
{Ionization plays a central role in the gas-phase chemistry of
  molecular clouds. Since ions are coupled with magnetic fields, 
  which can in turn counteract gravitational
  collapse, it is of paramount importance to measure their abundance
  in star-forming regions.
}
{We use spectral line observations of the high-mass star-forming region NGC 6334 I to 
derive the abundance of two of the most abundant molecular ions, HCO$^+$ and N$_2$H$^+$, and  consequently, the cosmic ray ionization rate. In addition, the line profiles provide information about the kinematics of this region.}
{We present high-resolution spectral line observations conducted with the HIFI instrument on board 
the~{\it Herschel Space Observatory} 
of the rotational transitions with {\textit J}$_{\rm up}\geq 5$ of the molecular species C$^{17}$O, C$^{18}$O, HCO$^+$, H$^{13}$CO$^+$, and N$_2$H$^+$.}
{The HCO$^+$ and N$_2$H$^+$ line profiles display a redshifted asymmetry consistent with a region of expanding gas. We identify two emission components in the spectra, each with a different excitation, associated with the envelope of NGC 6334 I. The physical parameters obtained for the envelope are in agreement with previous models of the radial structure of NGC 6334 I based on submillimeter continuum observations. 
Based on our new {\it Herschel}/HIFI observations, combined with the predictions from a chemical model, we derive a cosmic ray ionization rate that is an order of magnitude higher than the canonical value of $10^{-17}$ s$^{-1}$.}
{We find evidence of an expansion of the envelope surrounding the hot core of NGC 6334 I, which is mainly driven by thermal pressure from the hot ionized gas in the region. The ionization rate seems to be dominated by cosmic rays originating from outside the source, although X-ray emission from the NGC 6334 I core could contribute to the ionization in the inner part of the envelope.}


\titlerunning{Ionization toward the high-mass star-forming region 
NGC~6334~I} \authorrunning{J. L. Morales Ortiz et al.}

\maketitle

\section{Introduction}
\label{sec:intro}

High-mass stars form inside molecular clouds, in the densest and
coldest regions, and usually do so in groups (e.g., \citealp{zinnecker2007,
rivera-ingraham2013}). 
Given the involved timescales and luminosities, the original
molecular clouds where high-mass stars form are chemically and morphologically 
altered, and even disrupted by the energetic stellar winds and outflows, X-rays and
UV photons from the young stars, and $\ga$ MeV particles from the
explosions of supernovae in their vicinities. The cosmic rays (CRs) from supernovae 
are then diffused by magnetic fields in the galaxy joining the CRs
that pervade it, but before the information on their original site is
lost because of diffusion, CRs can be detected in the vicinity 
where they are accelerated by supernova shocks
because they ionize the gas more effectively than UV
photons and X-rays (e.g., \citealp{ceccarelli2011}). Specifically,
they ionize the H atoms and H$_2$ molecules of the Interstellar
Medium (ISM), creating H$_3^+$ ions. Since ion-neutral reactions are
much faster than neutral-neutral reactions in cold ($\leq 100$ K) gas
(as the latter usually have activation barriers), the CR ionization is
the starting point of the synthesis of the hundreds of molecules found in
cold molecular clouds, or, in other words, of their chemical
enrichment. 

The first molecular ions created from H$_3^+$ are HCO$^+$ 
and N$_2$H$^+$, by reactions with CO and N$_2$, respectively 
(e.g., \citealp{turner1995}). These ions can, therefore, be used 
to measure the CR ionization rate. For the reasons 
mentioned above, this is particularly interesting in high-mass
star-forming regions, because it provides us with crucial information
on the CR formation and, possibly, how it influences the star
formation process. Another crucial, but poorly constrained, aspect is
how deeply CRs can penetrate into a molecular cloud (e.g., 
\citealp{padovani2009}). This is predicted to have important consequences
in, for example, the accretion through protoplanetary disks and, consequently,
in the planet formation process (e.g., \citealp{balbus1998,gammie1996}). 
So far, only a few constraints exist on the penetration of CRs
at H$_2$ column densities higher than $\sim 10^{23}$\,cm$^{-2}$
\citep{padovani2013}. Such high column densities are reached in 
the envelopes of high-mass protostars and a few other sites of our Galaxy, 
so this is an additional reason for studying the CR ionization rate in high-mass 
star-forming regions.

In this work, we present new observations of the HCO$^+$ and
N$_2$H$^+$ molecular ions in the direction of a high-mass star-forming
region, NGC 6334 I (see Sect.~\ref{sec:source-background}). The
high-resolution spectral line observations were carried out with the
HIFI (Heterodyne Instrument for the Far Infrared) instrument on board
the {\it Herschel Space Observatory} as part of the CHESS
(Chemical HErschel Surveys of Star-forming regions) {\it Herschel} key
program \citep{ceccarelli2010}. The goal of the program is to obtain
unbiased line surveys of several star-forming regions (see, e.g.,
\citealp{ceccarelli2010}; \citealp{codella2010}; \citealp{zernickel2012};
\citealp{kama2013}).  In the survey of NGC 6334 I, we detect several
{\textit J}$_{\rm up}\geq 6$ HCO$^+$, H$^{13}$CO$^+$, and N$_2$H$^+$ 
transitions. Given the upper level energies $\geq$ 80 K and
critical densities $\geq 10^7$ cm$^{-3}$ of these transitions, the
CHESS observations allow us to probe the dense and warm gas in this
region and, consequently, the CR ionization rate deep inside the NGC
6334 I envelope. To better constrain the origin of the emission and
the physical conditions of the emitting gas, we also use the
C$^{18}$O and C$^{17}$O {\textit J}$_{\rm up}\geq 5$ transitions, detected
in the CHESS survey.

This paper is organized as follows: in Sect.~\ref{sec:src} we review
the main characteristics of NGC 6334 I and previous observations relevant
to this work. In Sect.~\ref{sec:obs} we describe the {\textit Herschel}/HIFI 
observations of the molecular tracers 
used in this paper. We then present the results of the analysis of the molecular line 
spectra in Sect.~\ref{sec:res}, describing the derived line parameters of the observed 
rotational transitions. In Sect.~\ref{sec:dis} we first discuss the origin of the observed 
line profiles and then, using a non-local thermodynamic equilibrium (LTE) large velocity gradient (LVG) radiative transfer code, we model the spectral line emission in order to estimate 
the physical parameters of the emitting gas (temperature, density, source size, and column 
density). Finally, we derive an estimate of the CR ionization rate across the region by 
comparing the measured HCO$^+$ and N$_2$H$^+$ relative abundances from the 
LVG analysis with those predicted by a chemical model. We summarize our conclusions 
in Sect.~\ref{sec:concl}.

\section{Source background}\label{sec:source-background}
\label{sec:src}

The source NGC 6334 I is a relatively close (1.7 kpc; \citealp{neckel1978})
high-mass star-forming region, with a total mass of $\sim$~200~M$_\odot$,
a bolometric luminosity of $\sim$\,2.6\,$\times 10^5$ L$_\odot$, and a size
of $\sim$ 0.1 pc, as determined from (sub)mm continuum observations 
\citep{sandell2000}. It is associated with an ultracompact HII region 
\citep{depree1995}, and masers have been detected in OH \citep{brooks2001}, 
H$_2$O \citep{migenes1999}, CH$_3$OH \citep{walsh1998}, and NH$_3$ 
\citep{walsh2007}. This source is a so-called hot core, since it possesses the 
main characteristics of these objects: high temperatures ($\gtrsim 100$ K), small 
sizes ($\lesssim 0.1$ pc), masses in the range $\sim 10 - 10^3$ M$_\odot$,
and luminosities larger than $10^4$ L$_\odot$ \citep{cesaroni2005}.

The NGC 6334 I region has been studied in considerable detail over the
last decades.  We summarize here some of the most recent results that
are relevant to the present work.  The Submillimeter Array (SMA) 1.3
mm continuum emission image toward NGC 6334 I (see Fig. 1 in
\citealp{hunter2006}) shows that the hot core itself consists of four
compact condensations located within a region $\approx10\arcsec $ in
diameter. These four sources are denoted I-SMA$1-4$ in descending
order of peak intensity. The I-SMA1 and I-SMA2 sources show rich
spectra of molecular transitions, I-SMA3 is associated with an HII
region excited by the NIR source IRS1E, and I-SMA4 shows dust emission
but no line emission \citep{zernickel2012}.

Spectral line observations of the NH$_3$(1,1) through (6,6) inversion
lines taken with the Australia Telescope Compact Array (ATCA;
\citealp{beuther2005} \& \citeyear{beuther2007}) show two emission
peaks associated with the two brightest continuum sources (I-SMA1 and
I-SMA2) reported by \cite{hunter2006}. The six inversion lines
observed by Beuther and co-authors cover a range of upper level
energies $E_{\rm up}$ from 23 K to 408 K. From the NH$_3$(5,5) and
(6,6) lines, \cite{beuther2007} assigned local standard of rest (LSR) 
velocities of $-7.6$ and $-8.1$\,km\,s$^{-1}$, and $-7.5$ and $-8.0$\,km\,s$^{-1}$ for 
the I-SMA1 and I-SMA2 sources, respectively. 

The radial structure of several star-forming cores, including NGC 6334
I, has been studied by \citet[hereafter R11]{rolffs2011} through
spectral line observations with the Atacama Pathfinder EXperiment
(APEX) and submillimeter continuum maps at 850 $\mu$m from the APEX
Telescope Large Area Survey of the GALaxy (ATLASGAL) taken with LABOCA
(Large APEX bolometer Camera).  By modeling the source as a centrally
heated sphere with a power-law density gradient, these authors first
reproduce the radial intensity profile of the continuum emission at
850 $\mu$m. Afterwards, they try to reproduce 
the spectral line emission of various HCN, HCO$^+$, and CO lines 
by radiative transfer modeling, 
providing the physical structure derived from the continuum analysis
as the input parameters. The models indicate that the density varies
with radius as~$r^{-1.5}$, which gives an effective dust temperature
of 65 K and a luminosity of 7.8$\, \times 10^4~$L$_\odot$ for NGC 6334
I. While the modeling of the radial profiles is consistent with the
continuum observations, the authors are only able to explain some of
the observed spectral features in the data. The
inconsistencies in the results are probably due to the simplified source
structure of the radiative transfer model and the actual complexity of
the internal structure of NGC 6334 I, which was already revealed by
the interferometric data from SMA.

Finally, previous CHESS observations of NGC 6334
I have already been reported by \cite{emprechtinger2010,
emprechtinger2013}, \cite{lis2010}, \cite{ossenkopf2010}, \cite{vanderwiel2010}, 
and \cite{zernickel2012}.
The first detection of H$_2$Cl$^+$, in the ISM, toward NGC 6334 I 
was reported by \cite{lis2010}. H$_2$Cl$^+$ is detected in absorption, and these 
authors report a column density in excess of 10$^{13}$ cm$^{-2}$. In addition, HCl is 
detected, in emission, at the hot core velocity ($\sim -6.3$ km\,s$^{-1}$). The CH 
emission component reported by \cite{vanderwiel2010} at a velocity of $-8.3$ km\,s$^{-1}$ 
is consistent with the H$_2$O hot core emission detected by \cite{emprechtinger2010} 
at $-8.2$\,km\,s$^{-1}$. The spectral line survey of NGC 6334 I by \cite{zernickel2012} 
identifies a total of 46 molecules, with 31 isotopologues, in a 
combination of emission and absorption lines. In particular, for the HCO$^+$ and 
H$^{13}$CO$^+$ lines the authors derive a velocity of $-7.5$ km\,s$^{-1}$ and a line width 
of 4.7 km\,s$^{-1}$, and a velocity of $-6.8$ km\,s$^{-1}$ and a line width of 4.6 km\,s$^{-1}$ 
for the CO isotopologues. Two components at velocities of $-6.8$ and $-9.5$ km\,s$^{-1}$ and 
line widths of 2.5 km\,s$^{-1}$ are determined for N$_2$H$^+$. 
The spectral analysis of water isotopologues by \cite{emprechtinger2013} 
identifies four physical components in the
line profiles. The first component, with a velocity of $-6.4$ km s$^{-1}$ 
and a line width of 5.4 km\,s$^{-1}$, corresponds to the
envelope of NGC 6334 I. The second component coincides with the 
LSR velocity of I-SMA2 ($\sim$\,\,$-8$ km\,s$^{-1}$), and
is thus associated with the embedded core. The other two components
are associated with foreground clouds and a bipolar
outflow. 

\section{Observations}
\label{sec:obs}

The observations were performed between February 28 and October 14, 2010, with the HIFI instrument on board the {\it Herschel Space Observatory} in the double beam switch mode 
(180$\arcsec$ chopper throw). The observed position with {\it Herschel}/HIFI toward NGC 6634 I 
($\alpha_{2000}= 17^h20^m53.32^s$, $\delta_{2000}= -35\degr46\arcmin58.5\arcsec$) 
is located between the I-SMA1 and I-SMA2 sources (Sect.~\ref{sec:src}). 
We employ the same data reduction process described in \cite{zernickel2012}; 
we briefly summarize here the data reduction steps. 
The data have been reduced with the HIPE (Herschel Interactive Processing Environment; 
\citealp{ott2010}) pipeline version 5.1. The spectral resolution of the double sideband (DSB) spectra, which were observed for 3.5 seconds each, is 1.1 MHz (corresponding to a 
velocity resolution ranging from 0.3 km\,s$^{-1}$ to 0.6 km\,s$^{-1}$). 
The DSB spectra have been observed with a redundancy of eight, which allows the 
deconvolution and isolation of the single sideband (SSB) spectra (\citealt*{comito2002}). 
The deconvolved SSB spectra have been exported to the FITS format for subsequent analysis using the GILDAS\footnote{http://www.iram.fr/IRAMFR/GILDAS} package. 
The half-power beamwidth (HPBW) and the main beam efficiency ($\eta_{\rm mb}$) for HIFI 
have been computed as a function of frequency for each of the observed lines by making linear interpolations from the values listed in \cite{roelfsema2012}. The $\eta_{\rm mb}$ values are then 
used to convert the intensities from the antenna temperature ($T_{\rm A}^\star$) to the main beam 
temperature ($T_{\rm mb}$) scale. 
The spectra shown in this paper are equally weighted averages of the horizontal 
and vertical polarizations. 
In the CHESS line survey we detect seven C$^{18}$O lines, 
five C$^{17}$O lines, 
seven HCO$^+$ lines, 
three H$^{13}$CO$^+$ lines, 
and four N$_2$H$^+$ lines. 
The spatial structure obtained from SMA (Sect.~\ref{sec:src}) is unresolved by {\it Herschel} because the beam covers the whole 
NGC 6334 I region (see Fig. 1 in \citealp{zernickel2012}). The SMA observations thus serve as an auxiliary tool 
to identify and separate the different components present in the HIFI spectra. 

\section{Results}
\label{sec:res}

\begin{figure}
\hspace{-0.5cm}
\includegraphics[width=7.0cm,angle=90]{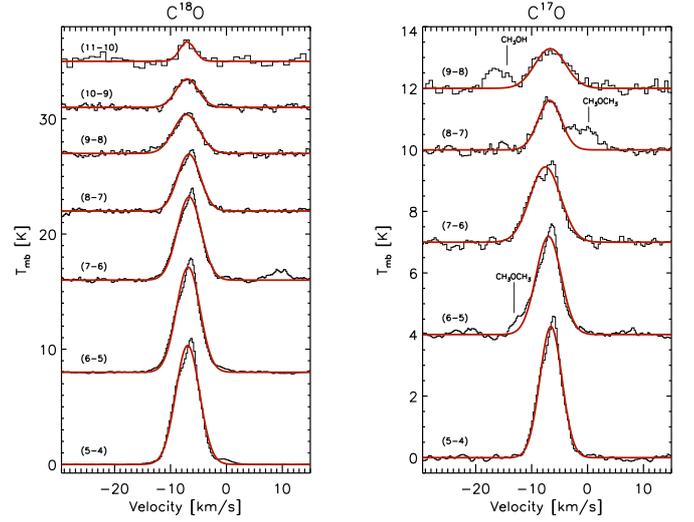}
\caption{Spectra of the C$^{18}$O and C$^{17}$O lines, baseline-subtracted and displaced with a vertical offset between each transition for clarity. The red lines show the Gaussian 
fits to the spectra. The vertical lines indicate the spectral features attributed to either CH$_3$OH or CH$_3$OCH$_3$.}
\label{fig:coSpectra}
\end{figure}

\begin{figure}
\hspace{-0.5cm}
\includegraphics[width=7.0cm,angle=90]{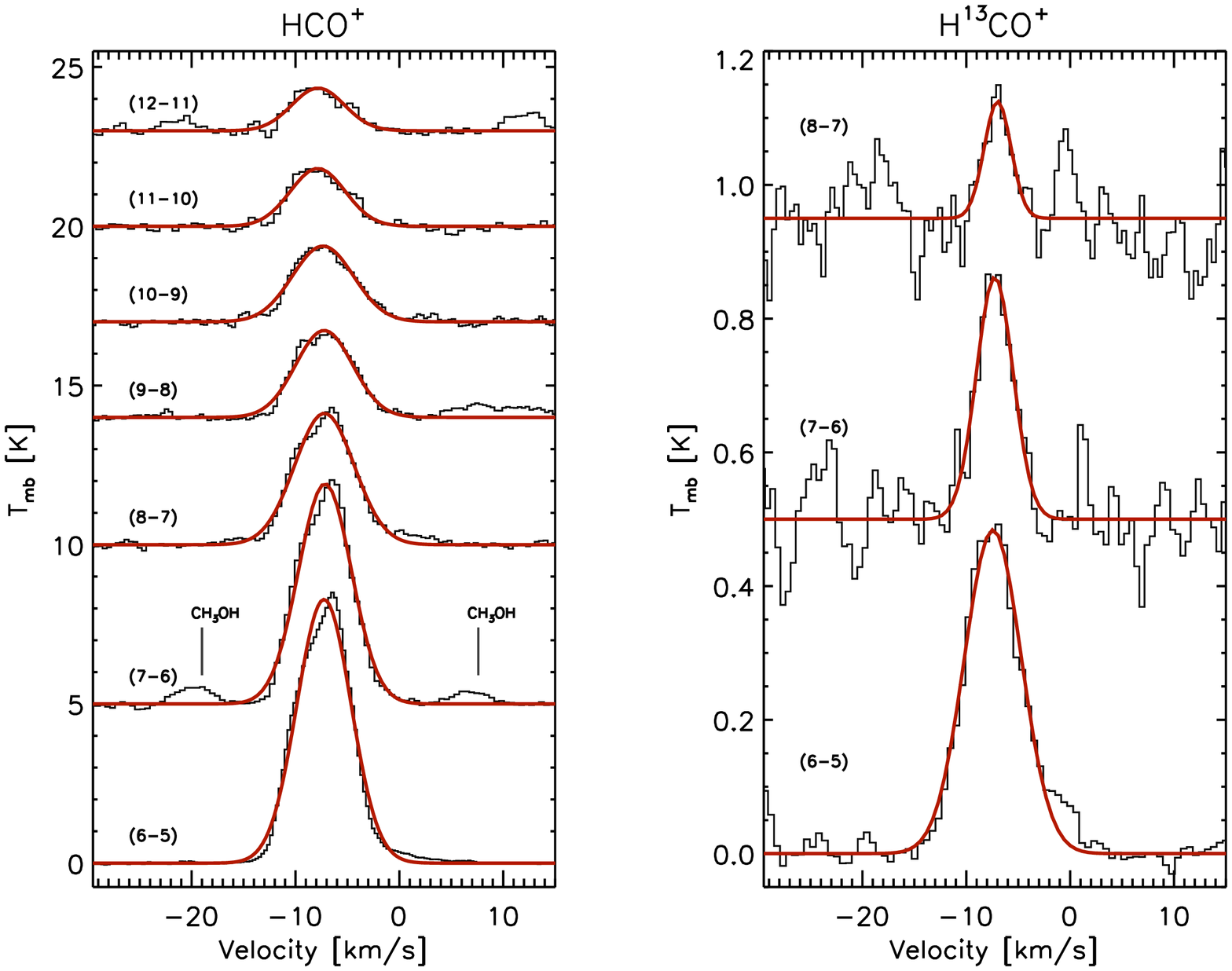}
\caption{Spectra of the HCO$^+$ and H$^{13}$CO$^+$ lines, baseline-subtracted and displaced with a vertical offset between each transition for clarity. The red lines show the Gaussian 
fits to the spectra. The vertical lines indicate the spectral features attributed to CH$_3$OH.}
\label{fig:hcoSpectra}
\end{figure}

\begin{figure}
\hspace{2.0cm}
\includegraphics[width=7.0cm,angle=90]{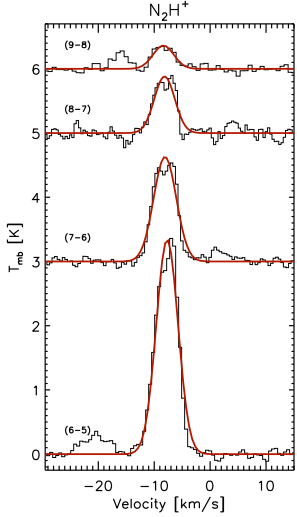}
\caption{Spectra of the N$_2$H$^+$ lines, baseline-subtracted and displaced with a vertical offset between each transition for clarity. The red lines show the Gaussian fits to the spectra.}
\label{fig:n2hSpectra}
\end{figure}

The spectral line data were analyzed using CLASS, which is part of the 
GILDAS\footnotemark[\value{footnote}] package. 
The spectra obtained with HIFI towards NGC 6334 I are shown in
Figs.~\ref{fig:coSpectra}$-$\ref{fig:n2hSpectra}. The line parameters, derived from 
a single Gaussian fit to the spectra, are listed in Table~\ref{tab:lines}. 
The RMS of the integrated intensities ($\sigma_{\rm I}$) in Table~\ref{tab:lines} is computed as 
$\sigma_{\rm I}=\sqrt{n \cdot (\sigma_{\rm base} \cdot \Delta \varv)^2+(f \cdot I)^2}$,
where $\sigma_{\rm base}$ is the baseline RMS, $\Delta \varv$ is the channel width, $n$ is the number of channels across the line profile that lie above the $\sigma_{\rm base}$ level of the spectrum, $I = \int{T_{\rm mb}\,{\rm d}V}$ is the integrated intensity; the factor {\it f}, which represents the calibration uncertainty of the HIFI instrument \citep{roelfsema2012}, is chosen to be 15\%. 
The rotational spectra of C$^{17}$O and N$_2$H$^+$ have a hyperfine structure. However, for the observed lines ({\textit J}$_{\rm up}\geq$ 5) the hyperfine components are so heavily blended together that, even when determined by a single Gaussian fit, the line width should not be significantly overestimated \citep{miettinen2012}. 

\begin{table*}
\caption{Line parameters.}
\label{tab:lines}
\centering
\begin{tabular}{lccccccc}
\hline\hline
Spectral Line &  Frequency & HPBW & $\eta_{\rm  mb}$ & $V_{\rm lsr}$ & $T_{\rm peak}$ & FWHM & $\int{T_{\rm mb}\,{\rm d}V}$ \\
 & [GHz] & [$\arcsec$] & & [km\,s$^{-1}$] & [K] & [km\,s$^{-1}$] & [K\,km\,s$^{-1}$] \\  
\hline
%
C$^{18}$O$(5-4)$   &  548.8437   & 39.5    & 0.76   & $-6.93 \pm 0.01$  & $10.3 \pm 0.1$ & $5.00 \pm 0.01$  &  $55.0 \pm 8$ \\
C$^{18}$O$(6-5)$   &  658.5683   & 32.4    & 0.75   & $-6.84 \pm 0.01$  &  $9.1 \pm 0.1$ & $4.93 \pm 0.01$  &  $48.0 \pm 8$ \\
C$^{18}$O$(7-6)$   &  768.2687   & 27.8    & 0.75   & $-6.67 \pm 0.03$  &  $7.2 \pm 0.3$ & $5.00 \pm 0.06$  &  $38.4 \pm 6$ \\
C$^{18}$O$(8-7)$   &  877.9416   & 24.4    & 0.75   & $-6.70 \pm 0.01$  &  $4.9 \pm 0.1$ & $4.78 \pm 0.03$  &  $25.1 \pm 4$ \\
C$^{18}$O$(9-8)$   &  987.5840   & 21.5    & 0.72   & $-7.16 \pm 0.04$  &  $3.4 \pm 0.2$ & $5.40 \pm 0.1$  &  $19.4 \pm 3$ \\
C$^{18}$O$(10-9)$   & 1097.1885   & 19.4    & 0.65   & $-7.02 \pm 0.06$  &  $2.5 \pm 0.2$ & $4.68 \pm 0.1$  &  $12.3 \pm 2$ \\
C$^{18}$O$(11-10)$   & 1206.7538   & 17.7    & 0.66   & $-7.02 \pm 0.3$  &  $1.8 \pm 0.4$ & $3.05 \pm 0.6$  &  $ 5.7 \pm 1$ \\

C$^{17}$O$(5-4)$   &  561.7251   & 38.6    & 0.75   & $-6.56 \pm 0.01$  &  $4.3 \pm 0.1$ & $4.64 \pm 0.03$  &  $21.0 \pm 3$ \\
C$^{17}$O$(6-5)$   &  674.0249   & 31.8    & 0.75   & $-6.97 \pm 0.02$  &  $3.2 \pm 0.1$ & $5.28 \pm 0.06$  &  $17.9 \pm 3$ \\
C$^{17}$O$(7-6)$   &  786.3005   & 27.1    & 0.75   & $-7.55 \pm 0.04$  &  $2.5 \pm 0.1$ & $6.40 \pm 0.1$  &  $16.7 \pm 3$ \\
C$^{17}$O$(8-7)$   &  898.5431   & 23.8    & 0.74   & $-6.77 \pm 0.06$  &  $1.6 \pm 0.1$ & $4.92 \pm 0.2$  &  $ 8.4 \pm 1$ \\
C$^{17}$O$(9-8)$   & 1010.7540   & 21.1    & 0.71   & $-6.67 \pm 0.3$  &  $1.3 \pm 0.2$ & $6.41 \pm 0.03$  &  $ 8.8 \pm 1$ \\

HCO$^+(6-5)$   &  535.0745   & 40.4    & 0.76   & $-7.25 \pm 0.02$  &  $8.3 \pm 0.1$ & $6.10 \pm 0.01$  &  $53.8 \pm 8$ \\
HCO$^+(7-6)$   &  624.2232   & 34.3    & 0.75   & $-7.16 \pm 0.03$  &  $6.9 \pm 0.2$ & $6.00 \pm 0.07$  &  $44.2 \pm 7$ \\ 
HCO$^+(8-7)$   &  713.3584   & 30.1    & 0.75   & $-7.22 \pm 0.02$  &  $4.1 \pm 0.1$ & $6.73 \pm 0.05$  &  $29.6 \pm 4$ \\
HCO$^+(9-8)$   &  802.4776   & 26.4    & 0.75   & $-7.24 \pm 0.07$  &  $2.7 \pm 0.2$ & $6.45 \pm 0.2$  &  $18.7 \pm 3$ \\
HCO$^+(10-9)$   &  891.5792   & 24.0    & 0.74   & $-7.36 \pm 0.04$  &  $2.4 \pm 0.1$ & $6.92 \pm 0.09$  &  $17.5 \pm 3$ \\
HCO$^+(11-10)$   &  980.6622   & 21.7    & 0.73   & $-7.86 \pm 0.08$  &  $1.8 \pm 0.1$ & $6.20 \pm 0.2$  &  $12.0 \pm 2$ \\
HCO$^+(12-11)$   & 1069.7219   & 19.9    & 0.67   & $-7.82 \pm 0.2$  &  $1.3 \pm 0.2$ & $5.94 \pm 0.4$  &  $ 8.5 \pm 1$ \\

H$^{13}$CO$^+(6-5)$   &  520.4728   & 41.5    & 0.76   & $-7.44 \pm 0.04$  &  $0.5 \pm 0.1$ & $6.44 \pm 0.1$  &  $ 3.3 \pm 0.5$ \\
H$^{13}$CO$^+(7-6)$   &  607.1893   & 35.5    & 0.75   & $-7.26 \pm 0.1$  &  $0.4 \pm 0.1$ & $4.17 \pm 0.3$  &  $ 1.6 \pm 0.3$ \\
H$^{13}$CO$^+(8-7)$   &  693.8924   & 30.9    & 0.75   & $-6.98 \pm 0.3$  &  $0.2 \pm 0.1$ & $3.04 \pm 0.6$  &  $ 0.6 \pm 0.1$ \\

N$_2$H$^+(6-5)$   &  558.9808   & 38.8    & 0.76   & $-7.69 \pm 0.06$  &  $3.3 \pm 0.2$ & $4.58 \pm 0.1$  &  $16.3 \pm 3$ \\
N$_2$H$^+(7-6)$   &  652.1131   & 32.7    & 0.75   & $-8.07 \pm 0.04$  &  $1.6 \pm 0.1$ & $4.76 \pm 0.09$  &  $ 8.3 \pm 1$ \\
N$_2$H$^+(8-7)$   &  745.2301   & 28.8    & 0.75   & $-8.15 \pm 0.09$  &  $0.9 \pm 0.1$ & $4.30 \pm 0.2$  &  $ 4.0 \pm 0.6$ \\
N$_2$H$^+(9-8)$   &  838.3306   & 25.5    & 0.75   & $-8.36 \pm 0.3$  &  $0.4 \pm 0.1$ & $4.37 \pm 0.6$  &  $ 1.7 \pm 0.3$ \\
\hline
\end{tabular}
\tablefoot{For each transition, we report the half-power beamwidth (HPBW), main beam efficiency 
($\eta_{\rm mb}$), LSR velocity ($V_{\rm lsr}$), peak intensity ($T_{\rm peak}$), full-width half-maximum
(FWHM), and integrated intensity ($\int{T_{\rm mb}\,{\rm d}V}$). All intensity values are in 
units of $T_{\rm mb}$ [K]. The peak intensity error only includes the statistical error. The integrated intensity error includes the statistical and calibration errors (see Sect.~\ref{sec:res}).}
\end{table*}

\begin{figure}
\hspace{-0.5cm}
\includegraphics[width=9.5cm,angle=0]{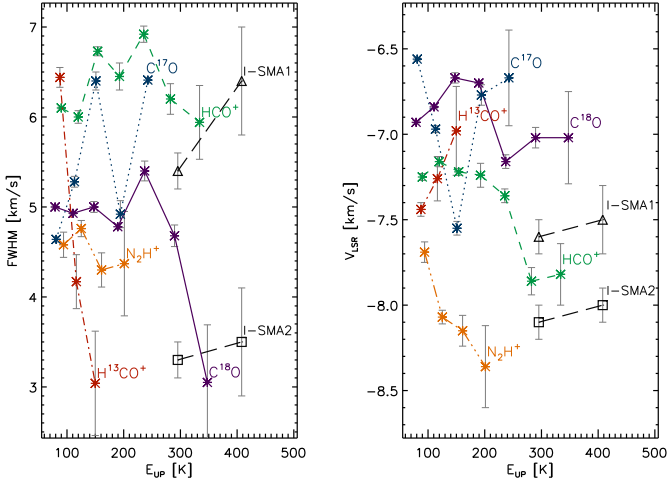}
\caption{FWHM (left panel) and velocity (right panel) vs. upper level energy {\textit E}$_{\rm up}$, for each of the molecular tracers. The NH$_3$(5,5) and (6,6) observations reported by \cite{beuther2007} for I-SMA1(triangles) and I-SMA2 (squares) are plotted as reference.}
\label{fig:EupFwhmVlsr}
\end{figure}

The observed molecular lines cover upper level energies in the range 79 K $\leq E_{\rm up} \leq$ 348 K, thus tracing different physical components within the NGC 6334 I region. 
The detected C$^{18}$O and HCO$^+$ lines have similar peak intensities, 
greater than the peak intensities of the N$_2$H$^+$ lines by a factor 
ranging from $\sim2$ to 10. Also, the average HCO$^+$/H$^{13}$CO$^+$ line intensity ratio 
is 18, which is lower than what would be expected in the case of optically thin line emission ($^{12}$C$/^{13}$C $\approx 75$; \citealp{langer1990}) by a factor of $\sim4$, thus indicating that the HCO$^+$ lines are optically thick. 
The C$^{18}$O lines also suffer from optical depth effects with an average C$^{18}$O/C$^{17}$O ratio of 2.8, which is lower than the expected value in the case of optically thin line emission  ($^{18}$O/$^{17}$O $\approx 3.5$; \citealp{penzias1981}). 
The C$^{18}$O and C$^{17}$O lines have an average value of
 $\langle V_{\rm lsr} \rangle =-6.9 \pm 0.3$ km\,s$^{-1}$, and the HCO$^+$ and 
H$^{13}$CO$^+$ lines have a value 
$\langle V_{\rm lsr} \rangle = -7.4 \pm 0.3$ km\,s$^{-1}$, 
while the N$_2$H$^+$ lines seem to have a slightly lower value of 
$\langle V_{\rm lsr} \rangle = -8.1 \pm 0.3$ km\,s$^{-1}$. 
These trends are shown in Figure~\ref{fig:EupFwhmVlsr}, which is a plot of the velocity and the full-width half-maximum (FWHM) as a function of the upper level energy ($E_{\rm up}$) for all the spectral line transitions. Taking into account all of the detected lines, we obtain $\langle V_{\rm lsr} \rangle = -7.3 \pm 0.5$ km\,s$^{-1}$. As a reference, we include in Figure~\ref{fig:EupFwhmVlsr} the ammonia observations reported by \cite{beuther2007} associated with the I-SMA1 and I-SMA2 sources. 

In terms of the line width, the HCO$^+$ lines are somewhat broader than the C$^{18}$O and 
N$_2$H$^+$ lines, and the C$^{17}$O lines seem to be divided between these two regimes 
(see Fig.~\ref{fig:EupFwhmVlsr}). 
The broadening of the HCO$^+$ lines, when compared to N$_2$H$^+$, 
is consistent with HCO$^+$ being optically thick, whereas N$_2$H$^+$ 
is an optically thin tracer. The spread in the C$^{17}$O line widths can be 
attributed to line contamination 
from other molecules, such as methanol (CH$_3$OH) and dimethyl ether (CH$_3$OCH$_3$), present in the C$^{17}$O spectra. The presence of methanol can also explain some of the spectral features visible in the HCO$^+$ spectra (see the HCO$^+(7-6)$ spectrum in Fig.~\ref{fig:hcoSpectra}, for example).
For the HCO$^+$ lines we have an average value of $\langle {\rm FWHM} \rangle = 6.3 \pm 0.4$ km\,s$^{-1}$, for the C$^{17}$O lines $\langle {\rm FWHM} \rangle = 5.5 \pm 0.8$ km\,s$^{-1}$, and for the C$^{18}$O, N$_2$H$^+$, and H$^{13}$CO$^+$ lines we have, respectively, 
$\langle {\rm FWHM} \rangle = 4.7 \pm 0.8$ km\,s$^{-1}$, 
$4.5 \pm 0.2$ km\,s$^{-1}$, and $4.6 \pm 1.7$ km\,s$^{-1}$. 
There is a spread between the H$^{13}$CO$^+$ lines that produces a larger uncertainty in the 
average FWHM values, which is probably due to the low signal-to-noise ratio (S/N) in 
the H$^{13}$CO$^+$ spectra, particularly in the $(7-6)$ and $(8-7)$ transitions. The 
H$^{13}$CO$^+(6-5)$ spectrum has the best S/N, and its FWHM is consistent with the 
line widths obtained from the HCO$^+$ lines. 
Taking into account all the lines, we obtain $\langle {\rm FWHM} \rangle =  5.2 \pm 1.1$ km\,s$^{-1}$. The average values obtained from our observations are in agreement, within the uncertainties, with the values obtained from the NH$_3$(5,5) and (6,6) observations by \citet{beuther2007}. In particular, the observed $V_{\rm lsr}$ and FWHM values seem to agree more with the parameters of the I-SMA1 source than with those of I-SMA2, but given that only two NH$_3$ transitions are available, and these occur at slightly higher energies compared to the other molecular transitions (see Fig.~\ref{fig:EupFwhmVlsr}), we cannot reach a definitive conclusion. In addition, the velocities 
($V_{\rm lsr}$ = $-7.7$ and $-7.6$ km\,s$^{-1}$) and line widths (FWHM = 5.3 and 5.1 km\,s$^{-1}$) obtained from the APEX observations \citep{rolffs2011} for the C$^{18}$O($6-5$) and ($8-7$) transitions agree with our results. 

\begin{figure}
\centering
\includegraphics[width=6.1cm,angle=90]{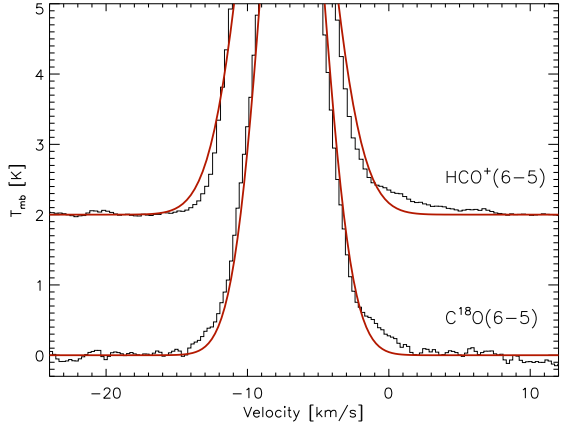}
\caption{Zoom in on the line-wing emission for the C$^{18}$O($6-5$) and HCO$^+(6-5)$ transitions. The red lines show the Gaussian fits to the spectra.}
\label{fig:outflows}
\end{figure}

In order to search for molecular outflows, we look for the presence of line-wing emission in the 
spectra and also in their residuals. Since the C$^{18}$O and HCO$^+$ lines are the brightest transitions, and do not have a hyperfine structure, they are the best candidates to search for outflows. We find that the line-wing emission is very minor (less than 4\% of the total integrated line intensity) and is most evident in the 
C$^{18}$O($6-5$) and HCO$^+(6-5)$ transitions, as shown in Figure~\ref{fig:outflows}. 
In particular, the line-wing emission in the HCO$^+(6-5)$ transition is only seen in the red-shifted side of the spectrum. Line-wing emission is also present in the C$^{18}$O$(5-4)$ and HCO$^+(8-7)$ transitions. In the higher-{\textit J} transitions, either there is no line-wing emission, or it is less evident than in the lower-{\textit J} transitions. 

\section{Discussion}
\label{sec:dis}

\subsection{Line profiles and kinematical components}
\label{sec:kin}

The spectral line profiles (Figs.~\ref{fig:coSpectra}$-$\ref{fig:n2hSpectra}) 
for the lower energy transitions of 
C$^{18}$O, C$^{17}$O, HCO$^+$, and N$_2$H$^+$ are asymmetric, with the
redshifted side of the profile being stronger than the blueshifted
side. This red asymmetry is consistent with emission from optically
thick lines from expanding gas. We note that the asymmetry is not seen in
the optically thin H$^{13}$CO$^+$ spectra, and is much less evident in
the higher energy transitions of the other molecules. Although it may
be just a fortuitous case, the simultaneous presence of the asymmetry
in the optically thick lines and absence in the optically thin lines
is evidence of the presence of an expanding motion of the
emitting gas \citep{mardones1997}. 

In the following, we present a quantitative analysis of the line profiles in support 
of the expanding gas hypothesis.  As mentioned above, in most cases, a single Gaussian 
does not fit the observations properly. We attempt to improve the analysis by using 
a double Gaussian fit in order to decompose the spectra into two kinematical components, 
which can be justified by the presence of two different velocity components in the line profile. 
Based on our analysis, and also on the results from previous observations toward NGC 6334 I, 
we initially assign a velocity of $-6.6$ km\,s$^{-1}$ to the emission from the envelope, and 
a velocity of $-9.8$ km\,s$^{-1}$ to the emission from the core. However, this fit is not unique, 
as demonstrated by an {\it a posteriori} analysis of the spectra, in which we used the results
from our radiative transfer analysis (Sect.~\ref{sec:lte}) as input parameters for a two-component Gaussian model. 
The spectra suggest the presence of a velocity gradient, which 
is also consistent with our hypothesis of an expanding envelope, as discussed in Section~\ref{sec:phys}. 

\begin{figure}
\centering
\includegraphics[width=8.8cm,angle=90]{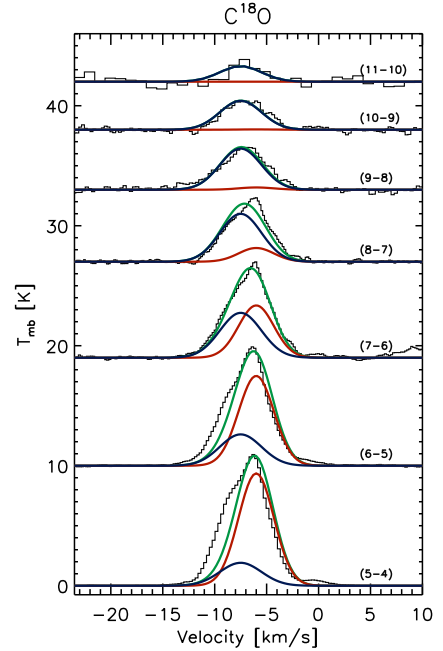}
\caption{Example of the C$^{18}$O spectra, overlaid with the double Gaussian profile derived from the results of the LVG analysis. The red, blue, and green lines show, respectively, the Gaussian profiles for the outer and inner envelope emission, and the sum of the two components.}
\label{fig:coSpectraLVG}
\end{figure}

As an example, we show in Figure~\ref{fig:coSpectraLVG} the double
Gaussian profile for the C$^{18}$O spectra. We assign, for the
outer and inner envelope emission respectively, velocities of $-6.0$ km\,s$^{-1}$ 
and $-7.5$ km\,s$^{-1}$, line widths of 4.0 km\,s$^{-1}$ and 4.7 km\,s$^{-1}$; the
line intensity ratios between the two components are the ones derived 
from the radiative transfer analysis. 
Since neither the single Gaussian fit nor the two-component Gaussian model can reproduce the line profiles accurately for all molecular transitions, we alternatively derive the integrated line intensities by computing the area obtained by summing all channels across the line profile and compare the results with the corresponding values obtained from the single Gaussian fit. Given that for the radiative transfer analysis we are mainly interested in the total emission from the spectra, and that the difference between the integrated intensities obtained by the two aforementioned methods ranges from 0$-$4\% (with a median value of 1\%), we conclude that the results from the single Gaussian profile are appropriate to perform the radiative transfer analysis. 

\subsection{Non-LTE analysis}
\label{sec:lte}

With the results obtained from the Gaussian fits
(Table~\ref{tab:lines}), and using a non-LTE LVG 
radiative transfer code \citep{ceccarelli2003}, we model the
spectral line emission of the observed transitions in order to obtain 
estimates of the temperature; H$_2$ density; size; and CO, HCO$^+$,
and N$_2$H$^+$ column density in NGC~6334~I. We achieve this by
comparing the observations with the LVG predictions and obtaining a
best-fit model by $\chi^2$ optimization. The optimization is done by
first finding the best reduced $\chi^2$ value for each input value of column
density, minimizing it with respect to the other model parameters
(i.e., temperature, molecular hydrogen density, and source size).
Afterwards, the minimum $\chi^2$ obtained from the set of column
densities corresponds to the model that best fits our results. The
uncertainties of the best-fit model are estimated by obtaining the 
adjustable parameter ranges within the 1-$\sigma$ confidence level, and 
using the appropriate number of degrees of freedom for each set 
of molecular transitions. The non-LTE LVG code makes use of the collisional
coefficients calculated by \cite{Flower1999} for HCO$^+$ and
N$_2$H$^+$. We note that \cite{Flower1999} reports calculations of the
HCO$^+-$ para-H$_2$ system for the first 21 energy levels in the
temperature range $5 \leq T \leq 390$ K. In our analysis, we use
the same coefficients for N$_2$H$^+$ and assume an ortho- to
para-H$_2$ ratio equal to 0.01, meaning that nearly all H$_2$ molecules are
in the para state. The collisional coefficients have been retrieved
from the BASECOL database\footnote{http://basecol.obspm.fr/} \citep{dubernet2013}.

\begin{figure}
\centering
\includegraphics[width=7.0cm,angle=0]{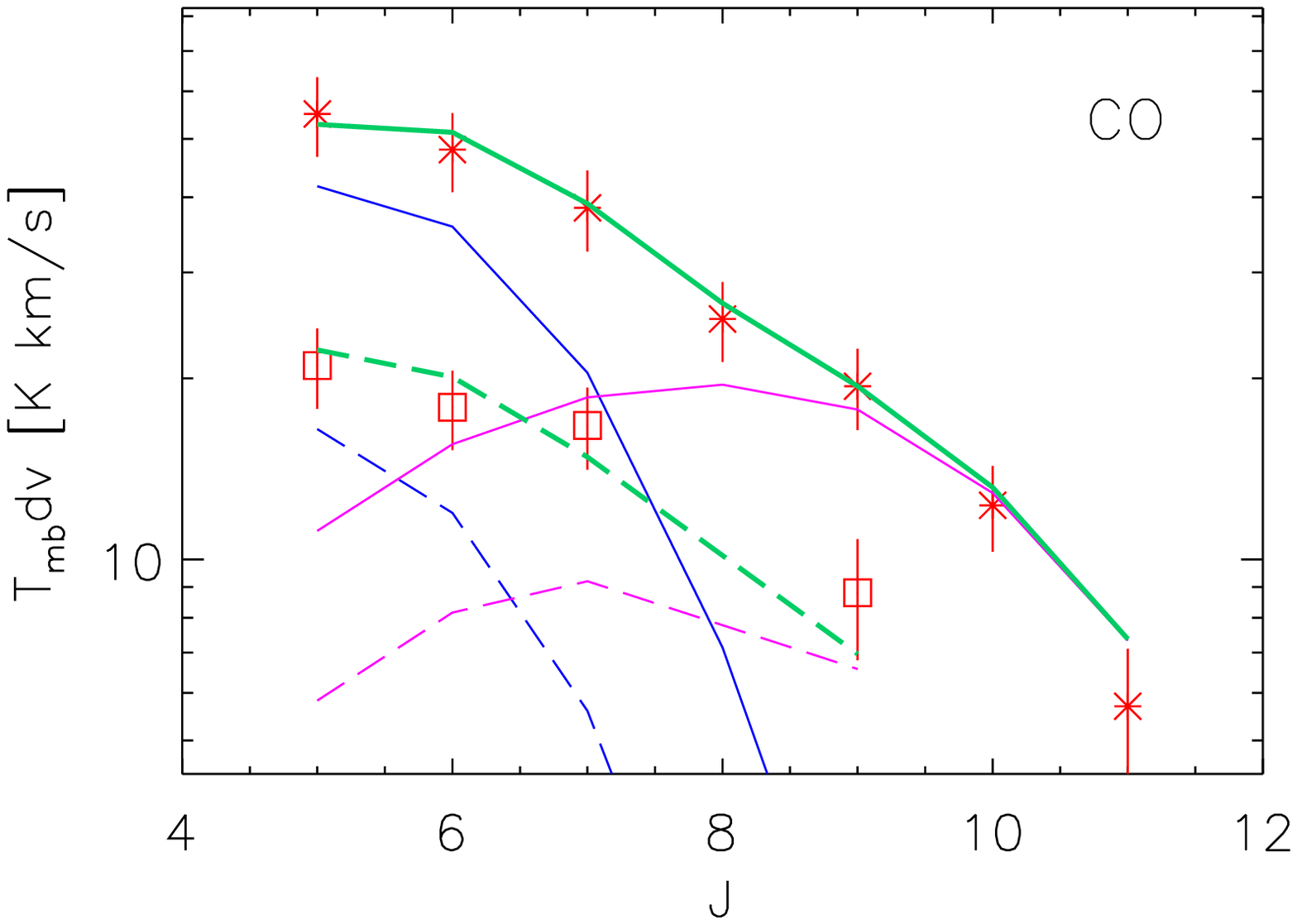}
\includegraphics[width=7.0cm,angle=0]{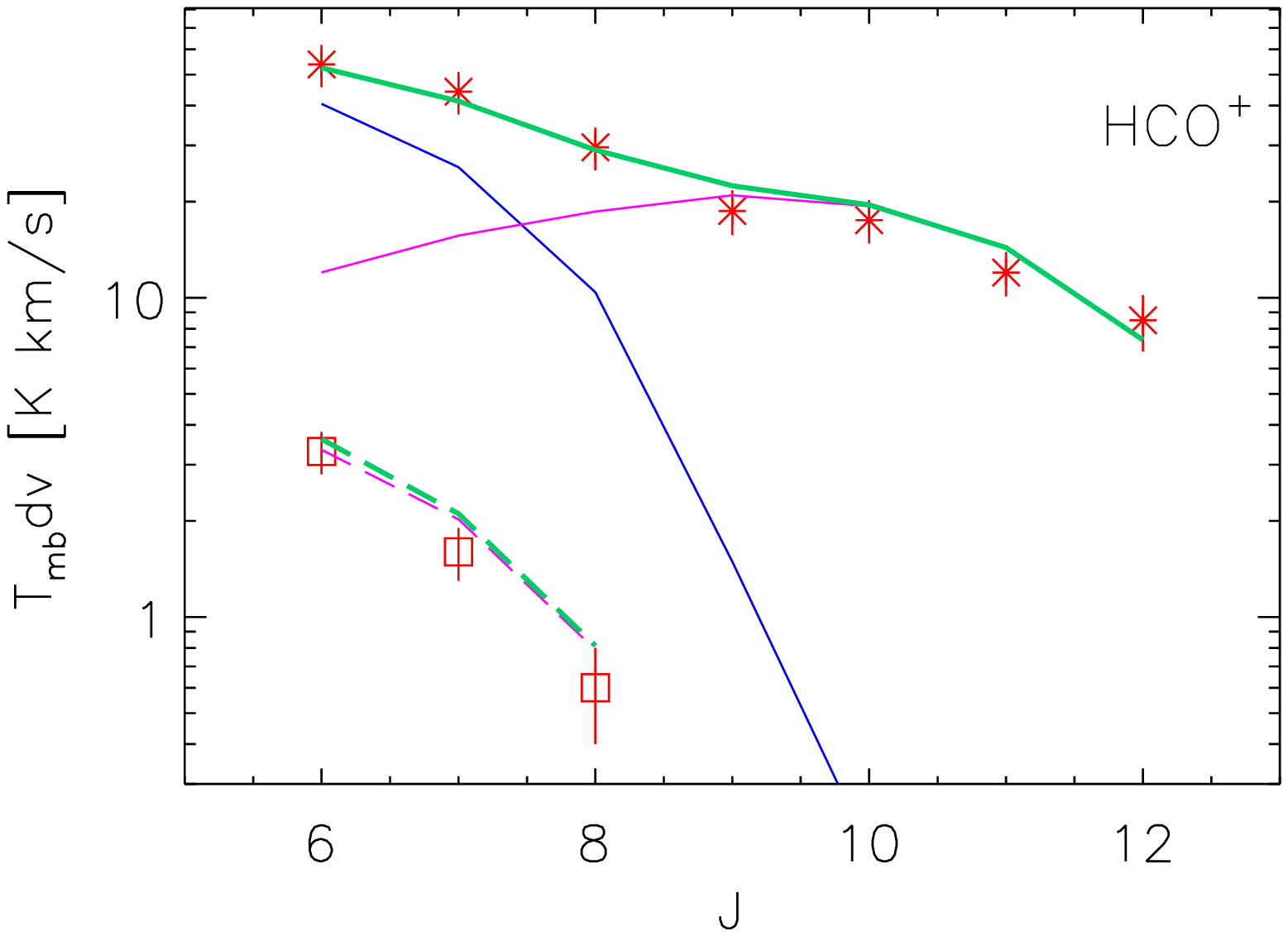}
\includegraphics[width=7.0cm,angle=0]{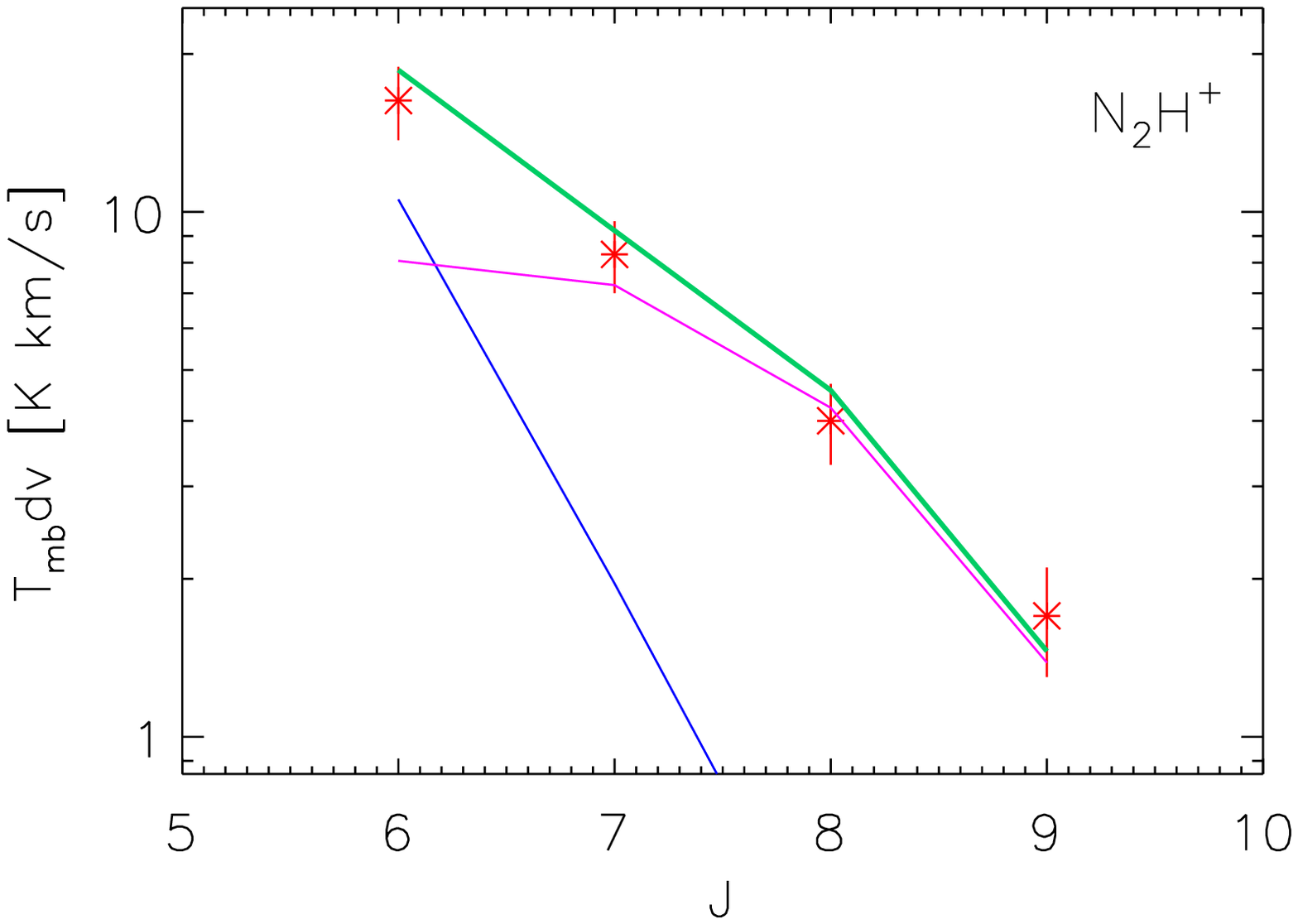}
\caption{Plots of integrated intensity vs. {\textit J}$_{\rm up}$ (from top to bottom) for the C$^{18}$O (red stars) and C$^{17}$O (red squares), HCO$^+$ (stars) and H$^{13}$CO$^+$ (squares), and N$_2$H$^+$ (stars) spectral line emission, obtained from the LVG model. The blue and pink lines represent, respectively, the theoretical model for the outer and inner envelope components, and the green lines represent the sum of the two components. The solid lines represent the model results for C$^{18}$O, HCO$^+$, and N$_2$H$^+$, while the dashed lines represent the model results for C$^{17}$O and H$^{13}$CO$^+$.}
\label{fig:lvg}
\end{figure}

The rotational transitions of HCO$^+$, H$^{13}$CO$^+$, N$_2$H$^+$, and
the CO isotopologues all have similar upper level energies and, with
the exception of the CO isotopologues, they all have similar critical
densities as well. All of these transitions are thus likely to probe the
same volume of gas within the observed region. Therefore, we aim to
find the model that best fits all the molecular transitions
simultaneously. To achieve this, we first fit the C$^{18}$O and
C$^{17}$O spectral line emission by $\chi^2$ optimization. Both
isotopologues are fitted simultaneously, thus imposing stronger
constraints on the model parameters. In practice, a one-component
model cannot fit the observations properly, indicating that there
are indeed two physical components with different physical conditions 
along the line of sight (Sect.~\ref{sec:kin}). We therefore fit the
observations with a two-component model:
{\it a posteriori}, the two components correspond to two regions of
the envelope (see Sect.~\ref{sec:phys}). Afterwards, the HCO$^+$ and
H$^{13}$CO$^+$ spectral line emission are fitted simultaneously,
using the H$_2$ density, temperature, and source size parameters
derived from the CO model as input values and adjusting the molecular 
column density in order to obtain the best fit.
This last step is repeated for the N$_2$H$^+$ spectral line emission,
so in the end all molecular transitions are fitted with a
two-component LVG model, in which the C$^{18}$O, HCO$^+$, and
N$_2$H$^+$ column densities are the only varying parameters. The
results from this method are presented in Table~\ref{tab:lvg} and
Figure~\ref{fig:lvg}.

\begin{table}
\caption{Results of LVG analysis.}
\label{tab:lvg}
\centering
\begin{tabular}{lcc}
\hline\hline
 & Outer envelope & Inner envelope \\
\hline
{\textit n}$_{\rm H_2}$ [cm$^{-3}$] & $(1.0 \pm 0.8) \times 10^5$ & $(1.0 \pm 0.5) \times 10^6$ \\
{\textit T} [K] & $35 \pm 4$ & $60 \pm 4$ \\
$\theta$ [$\arcsec$] & $40 \pm 6$ & $9 \pm 1$ \\
{\textit r} [AU] & $(3.4 \pm 0.5) \times 10^4$ & $(7.6 \pm 0.8) \times 10^3$ \\
{\textit N}[C$^{18}$O] [cm$^{-2}$] & $(6 \pm 1) \times 10^{16}$ & $(2 \pm 2) \times 10^{17}$ \\
{\textit N}[HCO$^+$] [cm$^{-2}$] & $(3 \pm 2) \times 10^{15}$ & $(7 \pm 1) \times 10^{15}$ \\
{\textit N}[N$_2$H$^+$] [cm$^{-2}$] &  $(5 \pm 2) \times 10^{14}$ & $(3 \pm 1) \times 10^{14}$ \\
\hline
\end{tabular}
\tablefoot{For each component, we report the H$_2$ density ({\textit n}), temperature ({\textit T}), angular diameter ($\theta$), linear radius ({\textit r}), and molecular column densities ({\textit N}). The column densities are averaged over the source size.}
\end{table}

We note that a small variation in the H$_2$ density, temperature, and source 
size values obtained from the CO model is necessary to best fit all molecular transitions simultaneously, thus making the radiative transfer modeling an iterative process. However, the 
variation between the integrated intensities of the $\chi^2$ optimized model and the final best-fit 
model is less than the uncertainties listed in the last column in Table 1.
We also note that the C$^{17}$O(8$-$7) transition is excluded from the LVG analysis because 
it is contaminated by CH$_3$OCH$_3$ emission (see Fig.~\ref{fig:coSpectra}). Finally, given the small number of detected N$_2$H$^+$ transitions, a one-component model could have 
been used to fit these observations, but we choose to consistently use the two-component 
model that is needed to explain the other molecular species considered in this work. 

\subsection{Origin of the emission: the expanding envelope}
\label{sec:phys}

\begin{figure}
\centering
\includegraphics[width=9.5cm,angle=0]{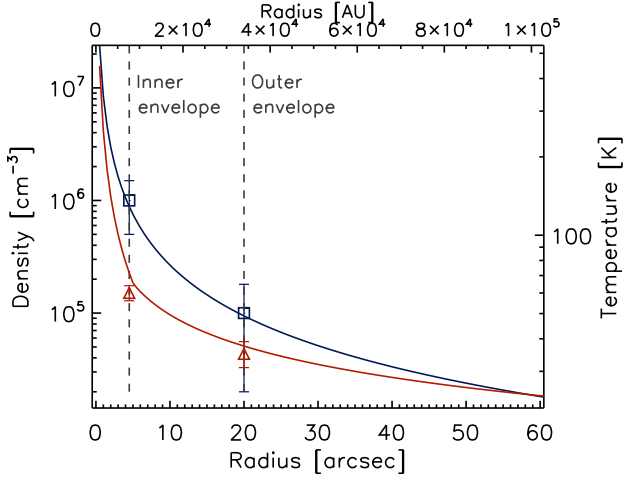}
\caption{Radial profiles of the density (blue line) and temperature (red line) toward NGC 6334 I from R11. The vertical lines indicate, respectively, the radius of the inner and outer regions obtained from our LVG analysis. The density (blue squares) and temperature (red triangles) values obtained for each component are overlaid on the plot.}
\label{fig:profiles}
\end{figure}

\begin{figure}
\hspace{-1.0cm}
\includegraphics[width=10.0cm,angle=0]{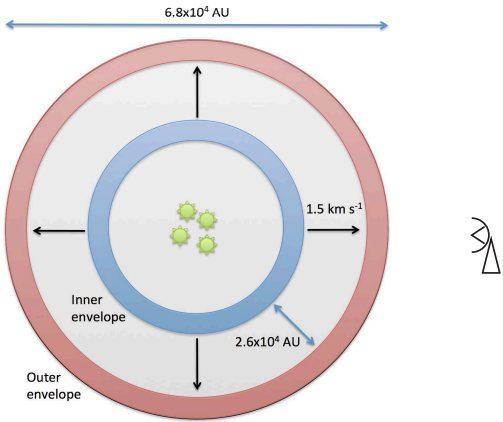}
\caption{Representation of the expanding envelope toward NGC 6334 I. See Sect.~\ref{sec:phys} for explanation.}
\label{fig:envelope}
\end{figure}

The radial profiles of density and temperature for NGC 6334 I,
obtained by using the specific parameters from the analysis of R11
(Sect.~\ref{sec:src}), are presented in Figure~\ref{fig:profiles}. The
parameters obtained from our LVG analysis (Table~\ref{tab:lvg}) are
overlaid on the profiles to compare the two methods. The
outer and inner envelope angular radii obtained from the LVG analysis,
which are also shown in Fig.~\ref{fig:profiles}, correspond 
to a density and temperature of $\sim \,$9$\, \times 10^4 \,$cm$^{-3}$ 
and 38 K for the outer region, and $\sim \,$9$\, \times 10^5 \,$cm$^{-3}$ 
and 72 K for the inner region. The results from our analysis
are thus remarkably consistent with the radial structure modeled by
R11 for NGC 6334 I.
As already mentioned in Section~\ref{sec:src}, the interferometric
observations from SMA \citep{hunter2006} revealed that the hot core of
NGC 6334 I is composed of various compact condensations within an
angular size of $\approx \,$10$\arcsec$. 
Although the {\it Herschel}/HIFI observations cannot resolve the
spatial structure of the hot core, the high level of agreement between
the results from our LVG analysis and the results from the radial
structure of R11 lead us to the conclusion that our spectral lines 
are dominated by the emission from the envelope of NGC 6334 I. 
We will therefore refer to the two components as the outer and inner envelope. 
In addition, as mentioned in Sect.~\ref{sec:kin}, the
red asymmetry displayed in the spectra is consistent with
expanding gas, and the Gaussian fits to the line profiles
suggest a velocity gradient in the envelope. The LVG
analysis shows that there are two major physical components that
we can identify with two kinematical components, with velocities of 
$-6.0$ km\,s$^{-1}$ and $-7.5$ km\,s$^{-1}$. 
We emphasize that these two velocity components are probably the two extremes 
of a gradual change in velocity through the envelope and we do not claim that our 
analysis fully reproduces the observed profiles. It makes, however, the important point 
that there is an expansion movement in the envelope with a gradient between the inner 
and outer envelope, namely on a scale of about 2.6 $\times 10^4$ AU, of about 1.5 km\,s$^{-1}$. 
The results are summarized in Figure~\ref{fig:envelope},
which is a representation of the expanding envelope. 
A full radiative transfer modeling of the line profiles as well as 
additional higher resolution spectral line observations to properly constrain the 
model parameters would be needed to determine the properties of 
the velocity gradient in the envelope, but this kind 
of analysis is beyond the scope of the present work. 

Previous studies of giant molecular clouds and HII regions (e.g.,
\citealp{bally1980, krumholz2009, lopez2011}) 
suggest that, given the expansion velocity and the radial distance at
which the expansion would be taking place, thermal pressure from hot ionized
gas should dominate the envelope expansion in NGC 6334 I. To give an
estimate, using Eq.~(9) in \citet{garcia1996} we find that the 
flux of ionizing photons produced by the UCHII region
associated with NGC 6334 I \citep{depree1995} can exert a thermal
pressure, {\textit P}$_{\rm t}$/k, on the surrounding envelope of 
1.6$\, \times 10^9$ cm$^{-3}$\,K. Using the results from our analysis, and 
the right-hand side of Eq.~(2) in \citet{olmi1999}, we estimate that 
the ambient pressure, {\textit P}$_{\rm a}$/k, exerted by the
envelope is between $10^8$ cm$^{-3}$\,K and $10^9$ cm$^{-3}$\,K.
Therefore, it is energetically possible to have an envelope expansion
in NGC 6334 I. The derived physical parameters for the expanding
envelope are discussed further in the following section.

We note that our conclusion of an expanding envelope toward NGC 6334 I
is subject to some {\it caveats}. First, the analysis of the SMA
data by \cite{zernickel2012}, as well as the ATCA observations by
\cite{beuther2007} (Sect.~\ref{sec:src}), indicate a
velocity difference between the I-SMA1 and I-SMA2 cores. The emission
from these two sources could potentially introduce line asymmetries
such as the ones observed in the HIFI spectra. Alternatively, the observed line 
profiles could be caused by emission from gas that is entrained in 
an outflow near or along the line of sight. 
However, we consider both these scenarios unlikely for the following reasons: 

\begin{enumerate}[(i)]
\item The HIFI observations also sample gas at spatial scales that have been
filtered out in the SMA maps. Thus, the velocity difference we measure
may have a different origin compared to that between I-SMA1 and
I-SMA2. 

\item The abundance of molecular ions is expected to be
lower toward the very central region of NGC 6334 I because of the lower
CR ionization in such a dense region. Our HCO$^+$ and N$_2$H$^+$
observations are thus more likely to probe a much larger volume of gas
compared to that of the interferometric observations.

\item The HCO$^+$ and N$_2$H$^+$ emitting sizes
(Table~\ref{tab:lvg}) are larger than those measured for the hot cores, 
while they are consistent with emission from a larger
scale, such as that from the envelope.

\item R11 found infall signatures in HCN high-excitation lines,
while our HCO$^+$ line profiles show no signs of infall, which
suggests that our spectra are mostly sensitive to less dense gas. 

\item The direction of the bipolar outflow in NGC 6334 I is in the northeast-southwest 
direction (PA$\, \sim \,$46$\degr$), as first reported by CO and SiO observations 
\citep{bachiller1990} and later confirmed by subsequent observations (e.g., 
\citealp{kraemer1995, leurini2006, beuther2008}). In particular, \cite{leurini2006} reported
that the small overlap between the outflow lobes indicates that the inclination angle of the
bipolar outflow is closer to $90\degr$ than to $0\degr$. It is thus unlikely that the outflow 
is responsible for the asymmetries observed in the line profiles, as it is far enough from 
the line of sight. We cannot exclude the possibility of another unresolved outflow along the 
line of sight, but this is not yet supported by any available data. 

\end{enumerate} 
However, the many observations now becoming available are 
revealing the complexity of this and other similar regions showing multiple
kinematic signatures. Therefore, building a complete picture of the
structure of NGC 6334 I will require multi-line interferometric
observations with zero-spacing to be sensitive to both core and
envelope components.

\subsection{Ionization structure of the envelope}
\label{sec:chem}

\subsubsection{Column densities and molecular abundances}
\label{sec:colden}

Using the results from the LVG analysis we can derive estimates of the H$_2$ column density, as well as the relative abundances between the various molecular species. With the column densities of HCO$^+$ and N$_2$H$^+$ listed in Table~\ref{tab:lvg} we obtain an [HCO$^+]/[$N$_2$H$^+$] abundance ratio of 6 and 23 for the outer and inner envelope components, respectively. We derive the CO column densities using an abundance ratio of [CO$]/[$C$^{18}$O] = 500 \citep{wilson1994}, and from the H$_2$ densities and the linear size of the inner and outer envelope we compute the H$_2$ column densities. Afterwards, we obtain the relative abundances of HCO$^+$ and N$_2$H$^+$ with respect to H$_2$ as well as the [CO]$/$[H$_2$] abundance ratios. These results are listed in Table~\ref{tab:param} where, with the exception of the [HCO$^+]/[$N$_2$H$^+$] ratio, the uncertainties of the derived parameters are estimated using the standard method of error propagation. For the [HCO$^+]/[$N$_2$H$^+$] abundance ratio, the uncertainties are instead determined by analyzing the variation in the ratio of the resulting HCO$^+$ and N$_2$H$^+$ column densities for all LVG solutions within the 1-$\sigma$ confidence level (see Sect.~\ref{sec:lte}). During this procedure the two sets of 
[HCO$^+]/[$N$_2$H$^+$] ratios (one for the outer and one for the inner envelope) used to determine the uncertainties are computed from pairs of solutions with equal (within a 5\% tolerance) model parameters (i.e., H$_2$ density, temperature, and source size). We choose to apply this method instead of the standard error propagation because the latter would artificially increase the resulting uncertainties because the uncertainties of the [HCO$^+]/[$H$_2$] and [N$_2$H$^+]/[$H$_2$] ratios must take into account the variation of H$_2$ density, gas temperature, and source size within the 1-$\sigma$ confidence level for the LVG model. 

As a comparison, we also list in Table~\ref{tab:param} the H$_2$ column densities computed using the radial density profile from R11 and the same method described above, which are consistent with our results. Although there is no clear trend in the molecular abundances when the uncertainties are taken into account, if we consider the [HCO$^+]/[$N$_2$H$^+$] abundance ratios we find that it is definitively different in the inner envelope when compared to that in the outer envelope. 

For the computation of the CO column densities we assume, as is common practice, that the [CO]$/$[C$^{18}$O] ratio equals the [$^{16}$O$]/[^{18}$O] ratio. This oxygen isotopic ratio has been estimated for the ISM by \citet{smith2009} as a function of the source distance from the Galactic center. Using  Eq.~(4) in \citet{smith2009} with a Galactocentric distance of 6.8 kpc for NGC 6334 \citep{kraemer1998,kraemer2000} we obtain a [$^{16}$O$]/[^{18}$O] ratio of 488$\, \pm \,$68 for NGC 6334 I. The difference with respect to the value of 500 adopted above for the [$^{16}$O$]/[^{18}$O] ratio is less than 3\%, and therefore does not affect our results for the relative abundances. 

For the [CO]$/$[H$_2$] abundance ratio, we obtain values of 3$\, \times 10^{-4}$ and 5$\, \times 10^{-4}$ for the outer and inner envelope, which are higher than the standard dense cloud value of $10^{-4}$ (\citealp{burgh2007}, and references therein). These abundance ratios are obtained using H$_2$ column densities of $10^{23}$ and 2$\, \times 10^{23}$ cm$^{-2}$ for the outer and inner envelope respectively, as listed in Table~\ref{tab:param}. There are two alternative methods of computing the H$_2$ column density in addition to the one previously described. 

The first alternative method is to use the derived CO column density and the [CO]$/$[H$_2$] standard dense cloud value to compute the H$_2$ column density. Using this method we obtain higher H$_2$ column densities by a factor of 3 and 5 for the outer and inner envelope of NGC 6334 I. The second alternative method is to assume the source has a spherical symmetry and a radial density profile of the form
\begin{equation}
n(r) = \frac{n_0}{1+(r/R)^{1.5}}\,,
\label{eq:nr}
\end{equation}
where $n_0$  and $R$ are, respectively, the H$_2$ density and envelope radius for the outer and inner components. The H$_2$ column densities are then computed by integrating this density profile along the line of sight, after a convolution with a 2D Gaussian corresponding to the HIFI beam. 
Using this method, we obtain H$_2$ column densities that are lower by a factor of 1.7 and 2, and [CO]$/$[H$_2$] abundance ratios of 5$\, \times 10^{-4}$ and $10^{-3}$ for the outer and inner envelope. The results from this method are within the uncertainties of the values listed in Table~\ref{tab:param}. A more detailed radiative transfer analysis that also takes into account the non-spherical envelope structure would be needed to better understand the variations in the column density and [CO]$/$[H$_2$] abundance ratio, but this analysis is beyond the scope of the present work. 

\begin{table}
\caption{Derived parameters.}
\label{tab:param}
\centering
\begin{tabular}{lcc}
\hline\hline
 & Outer envelope & Inner envelope \\
\hline
{\textit N}[H$_2$] [cm$^{-2}$]\tablefootmark{a} & $(1.0 \pm 0.8) \times 10^{23}$ & $(2 \pm 1) \times 10^{23}$ \\
$[$HCO$^+]/$[N$_2$H$^+$] & $6 \pm 2$ & $23 \pm 3$ \\
$[$CO]$/$[H$_2$] & $(3 \pm 2)\times 10^{-4}$ & $(5 \pm 5)\times 10^{-4}$ \\
$[$HCO$^+]/$[H$_2$] & $(3 \pm 3) \times 10^{-8}$ & $(3.5 \pm 2) \times 10^{-8}$ \\
$[$N$_2$H$^+]/$[H$_2$] & $(5 \pm 4)\times 10^{-9}$ & $(1.5 \pm 0.8)\times 10^{-9}$ \\
\hline
{\textit N}[H$_2$] [cm$^{-2}$]\tablefootmark{b} & $(9.6 \pm 2) \times 10^{22}$ & $(2.0 \pm 0.5) \times 10^{23}$ \\
\hline
$\zeta^{\rm H_2}$ [s$^{-1}$] & $(2.0\, \substack{+8\,\,\,\, \\ -1.3}) \times 10^{-16}$ & $(8.5\, \substack{+3.5 \\ -1.7}) \times 10^{-17}$ \\
\hline
\end{tabular}
\tablefoot{For each component, we give the H$_2$ column density ({\textit N}) averaged over the source size, relative abundances (with respect to H$_2$), and cosmic ray ionization rate ($\zeta^{\rm H_2}$). \tablefoottext{a}{Computed using the results of LVG analysis (Sect.~\ref{sec:colden}).} \tablefoottext{b}{Computed using the results of the radial profiles from R11 (Sect.~\ref{sec:phys}).}}
\end{table}

\subsubsection{Chemical modeling}
\label{sec:crir}

We model the chemical evolution of the source using the gas-phase reaction network from the KInetic Database for Astrochemistry (KIDA) in conjunction with the Nahoon code \citep{wakelam2012}. The KIDA network contains over 6000 unique chemical reactions involving almost 500 different species and a total of 6467 rate coefficients. The Nahoon code uses this database to compute the time-dependent gas-phase chemistry at a fixed temperature (between 10$\,-\,$300 K), density, and CR ionization rate. A set of reactions involving grains that are not included in KIDA, such as the formation of H$_2$ on grains, are also included in Nahoon (see Table 3 in \citealp{wakelam2012} for a description). The formation of HCO$^+$ and N$_2$H$^+$ in dense regions (e.g., \citealp{turner1995,bergin1997a}) is dominated by the CR ionization rate. Consequently, the abundances of these molecular ions, 
as well as their abundance ratio, provide us with an ideal tool to estimate the CR ionization rate toward the envelope of NGC 6334 I. By running a $10 \times 10$ grid of Nahoon models, sampled logarithmically in a range of H$_2$ densities and CR ionization rates ($\zeta^{\rm H_2}$), we compute the expected [HCO$^+$]/[N$_2$H$^+$] abundance ratios for the two temperature components. In these models, the initial abundances are taken from Table 2 in \cite{wakelam2010} where all species are in the atomic form except for molecular hydrogen, and the final abundances are computed for steady state ($10^7$ yr). Through the comparison of the model results with the H$_2$ densities and [HCO$^+$]/[N$_2$H$^+$] abundance ratios derived from the LVG analysis, we then obtain an estimate for $\zeta^{\rm H_2}$ toward NGC 6334 I. 

\begin{figure}
\centering
\includegraphics[width=9.0cm,angle=0]{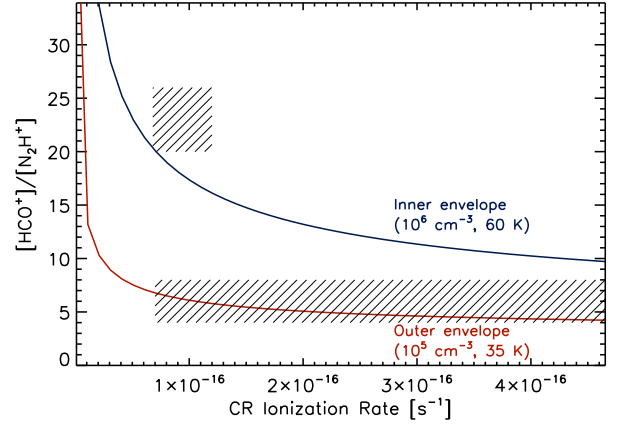}
\caption{[HCO$^+$]/[N$_2$H$^+$] abundance ratio as a function of the CR ionization rate computed for the inner (blue line) and outer (red line) envelope components using the Nahoon code. The shaded areas represent the results obtained within the uncertainties listed in Table~\ref{tab:param}. The $\zeta^{\rm H_2}$ values in the upper-left and lower-right corners of the shaded areas correspond to the maximum and minimum values of the [HCO$^+$]/[N$_2$H$^+$] ratio, respectively (see text for discussion).}
\label{fig:cr}
\end{figure}

Using this method we find, for the outer and inner envelope components respectively, CR ionization rates of 2.0$\, \times 10^{-16}$~s$^{-1}$ and 8.5$\, \times 10^{-17}$ s$^{-1}$, which are both higher than the standard CR ionization rate in molecular clouds of $\zeta^{\rm H_2}\approx10^{-17}$ s$^{-1}$ (\citealp{padovani2009}, and references therein). We note that the values of $\zeta^{\rm H_2}$, and also the H$_2$ column densities, obtained for NGC 6334 I are similar to previous values obtained by \citet{vandertak2000} and \citet{doty2002} toward the envelopes surrounding massive protostellar sources. In Figure~\ref{fig:cr} we present the computed [HCO$^+$]/[N$_2$H$^+$] abundance ratio as a function of the CR ionization rate for the outer and inner envelope components, in which the H$_2$ densities and temperatures obtained from the LVG analysis have been used as input parameters for the Nahoon model. Overlaid are the ranges of abundance ratios and CR ionization rates obtained for each component when the [HCO$^+$]/[N$_2$H$^+$] ratio is varied 
by the uncertainties listed in Table~\ref{tab:param}. There seems to be a trend for the CR ionization rate in the envelope to increase outward, but because of our uncertainties (see Fig.~\ref{fig:cr}) this conclusion remains uncertain. 

The tentative trend for the CR ionization rate would suggest that the main source of ionization originates outside of NGC 6334 I. It has already been mentioned that the ionization in dense regions, like the envelope of NGC 6334 I, is dominated by CRs. But X-ray emission from young massive stars in the surrounding molecular cloud could also contribute as a source of ionization. X-ray emission in the NGC 6334 giant molecular cloud has been revealed through observations with the {\it Advanced Satellite for Cosmology and Astrophysics} (ASCA; \citealp{sekimoto2000}) and the {\it Chandra X-Ray Observatory} \citep{ezoe2006} in the energy range of 0.5$-$10 keV. The X-ray emission is mainly associated with young massive stars embedded in the five far-infrared (FIR) regions denoted NGC 6334 I$-$V. Therefore, the nearest sources of X-rays should be cores II$-$V in NGC 6334 and the interior of NGC 6334 I itself. From the ASCA observations, \citet{sekimoto2000} obtained X-ray luminosities ($L_{\rm X}$) and hydrogen column densities ($N_{\rm H}$) for each of the five cores, and then computed the radius ($r_{\rm X}$) inside which the X-ray ionization rate ($\zeta_{\rm X}$) is comparable to the standard value for the CR ionization rate using the equation given by \cite{maloney1996}: 
\begin{equation}
\zeta_{\rm X} = 1.4 \times 10^{-18} {\rm \,s}^{-1} \frac{L_{\rm X}} {10^{33} \rm \,erg\,s^{-1}} \frac {1 \rm \,pc^2} {r_{\rm X}^2} \frac {10^{22} \rm \,cm^{-2}} {N_{\rm H}}\,.
\label{eq:Zx}
\end{equation}
With X-ray luminosities of $\sim \,$10$^{33}$ erg\,s$^{-1}$ and column densities of $\sim \,$10$^{22}$ cm$^{-2}$ for each of the five cores, and assuming $\zeta_{\rm X} = 10^{-17}$ s$^{-1}$, Sekimoto and co-authors determined that the X-ray ionization rate is comparable to that by CRs within a radius of $\sim 0.3$ pc from each core. This distance is much smaller than the distance between cores II$-$V and core I, and therefore there should be no significant ionization in NGC 6334 I due to external X-ray sources. The authors obtained $L_{\rm X} = 0.66 \times 10^{33}$\,erg\,s$^{-1}$ and $N_{\rm H} = 2.0 \times 10^{22}$ cm$^{-2}$ for the X-ray emission from the interior of NGC 6334 I. By using equation (\ref{eq:Zx}) with our $\zeta^{\rm H_2}$ values from Table~\ref{tab:param}, we obtain that X-ray ionization for the outer envelope ($r = 0.16$ pc) is comparable to CR ionization up to a radius of 0.05 pc. The ionization in the outer envelope should thus be dominated by CRs. However for the inner envelope ($r = 0.04$ pc) X-ray ionization is comparable to that by CRs up to a radius of 0.07 pc. Therefore the CR ionization rate value for the inner envelope should be treated as an upper limit due to any additional contribution from X-ray emission. 

Finally, we compute the theoretical value of the CR ionization rate as a function of the column density of traversed matter for NGC 6334 I using the model from \cite{padovani2009}. The model provides four possible profiles of the CR ionization rate depending on how the energy distribution of CR particles incident on the source is defined, and includes a detailed treatment of CR propagation that  takes into account the decrease of the ionization rate as the CRs penetrate the cloud. Using the {\textit N}[H$_2$] column densities in Table~\ref{tab:param} we find that the M02+E00 model\footnote{In this model, the energy distribution of CR particles is defined using the estimates of \cite{moskalenko2002} and \cite{strong2000}. See \cite{padovani2009} for a detailed description.} of \cite{padovani2009} provides the profile that best agrees with our results, and from it we obtain CR ionization rates of 1.4$\, \times 10^{-16}$ s$^{-1}$ and 9.5$\, \times 10^{-17}$ s$^{-1}$ for the outer and inner envelope components, respectively. 
The agreement, within the uncertainties, between the results from the KIDA/Nahoon model and the model from \cite{padovani2009} suggests that interstellar CRs alone are capable of producing the ionization rates obtained for NGC 6334 I and, therefore, additional sources of ionization are not required in order to explain our results. 

\section{Summary and conclusions }
\label{sec:concl}

We presented high-resolution spectral line observations of the
high-mass star-forming region NGC 6334 I obtained with the HIFI
instrument on board {\it Herschel}. In total, five molecular species were analyzed
through several rotational transitions with {\textit J}$_{\rm up}\geq5$. The
results of our study can be summarized as follows:

\begin{enumerate}
\item We detect bright emission from the HCO$^+$ and N$_2$H$^+$
  molecular ions, which allows us to constrain the ionization in the
  warm and dense regions of NGC 6334 I.

\item Optically thick HCO$^+$ and N$_2$H$^+$ lines display profiles
  with a redshifted asymmetry,
  while optically thin H$^{13}$CO$^+$ lines have a Gaussian profile peaked at the
  absorption dip of the optically thick lines. This is consistent with an expanding gas.

\item A non-LTE LVG radiative transfer analysis of the integrated line intensities shows the
  presence of two major physical components whose temperatures and densities
  are consistent with those previously derived from the radial structure of the
  envelope of NGC 6334 I.

\item We find evidence supporting the scenario in which the envelope surrounding the 
  hot core of NGC 6334 I is expanding and thermal pressure from hot ionized gas is able to 
  provide the driving force required for the envelope expansion.

\item We find a tentative trend for the CR ionization rate in the envelope
  to increase outward, which would suggest that the main source of ionization 
  lies outside NGC 6334 I. The ionization in the outer envelope is probably 
  dominated by CRs, but there could be an additional contribution to the
  ionization in the inner envelope from X-ray emission originating in
  the interior of NGC 6334 I.

\item The values obtained for the CR ionization rate are consistent,
  within the errors, with the model from \cite{padovani2009} that computes the CR
  ionization rate through a detailed treatment of CR propagation in
  molecular clouds, and they are also similar to previous values
  obtained toward the envelopes surrounding massive protostellar
  sources.

\end{enumerate}

\begin{acknowledgements}
HIFI has been designed and built by a consortium of institutes and university departments from across Europe, Canada and the United States under the leadership of SRON Netherlands Institute for Space Research, Groningen, The Netherlands, and with major contributions from Germany, France and the US. Consortium members are: Canada: CSA, U.Waterloo; France: CESR, LAB, LERMA, IRAM; Germany: KOSMA, MPIfR, MPS; Ireland, NUI Maynooth; Italy: ASI, IFSI-INAF, Osservatorio Astrofisico di Arcetri-INAF; Netherlands: SRON, TUD; Poland: CAMK, CBK; Spain: Observatorio Astron�mico Nacional (IGN), Centro de Astrobiolog�a (CSIC-INTA). Sweden: Chalmers University of Technology - MC2, RSS \& GARD; Onsala Space Observatory; Swedish National Space Board, Stockholm University - Stockholm Observatory; Switzerland: ETH Zurich, FHNW; USA: Caltech, JPL, NHSC. 
Support for this work was provided by NASA through an award issued by JPL/Caltech. 
JMO acknowledges the support of NASA, through the PR NASA Space Grant Doctoral Fellowship, and from the Institut de Plan\'etologie et d'Astrophysique de Grenoble (IPAG). C.Ceccarelli acknowledges the financial support from the French Agence Nationale pour la Recherche (ANR)
(project FORCOMS, contract ANR-08-BLAN-0225) and the French spatial agency CNES. 
The authors wish to thank Marco Padovani for providing the models for computing the CR ionization rate, as well as the anonymous referee and Ana L\'{o}pez-Sepulcre, whose comments much contributed to improve this work. 
\end{acknowledgements}


\bibliographystyle{aa}
\bibliography{refs}

\end{document}